# Diffractive effects in singly-resonant continuous-wave parametric oscillators

Kai Drühl

Center for Technology Research

Maharishi University of Management

Fairfield, IA 52557-1074

## Abstract

This paper presents a detailed numerical study of the effect of focusing on the conversion efficiency of low-loss singly-resonant parametric oscillators with collinear focusing of pump and signal. Results are given for the maximal pump depletion and for pump power levels required for various amounts of depletion, as functions of pump and signal confocal parameters, for $k_I/k_P=0.33$ and $0.50$. It is found that the ratio of pump depletion to maximal depletion as a function of the ratio of pump power to threshold power agrees with the plane-wave prediction to within 5%, for a wide range of focusing conditions. The observed trends are explained as resulting from intensity and phase dependent mechanisms.

## Introduction

The recent demonstrations of continuous-wave operation for singly-resonant optical parametric oscillators (SRO) with significant pump depletion in collinear focusing configurations [1-6] have stimulated renewed interest in these devices, and call for a more detailed analysis of diffractive effects. An exact calculation of thresholds for collinear focusing was presented in [7]. Theoretical studies of SROs above threshold so far have used the plane-wave [2,4,8] or lowest-order Gaussian mode [1,2,4,5,9] approximations. A numerical model of a pulsed SRO with strong birefringent walk-off is found in reference [10]. In [11], the author studied the effects of residual dispersion and diffraction on threshold, pump depletion and transverse mode content in a model of a low-loss SRO with collinear focusing at the center of the crystal, and a confocal parameter of signal and pump equal to the crystal length. It was found that the transverse mode content of the idler depends strongly on the residual dispersion, but only weakly on the amount of pump depletion. At optimal dispersion, about 77% of the total idler power is in the zero order radial mode, 22% in the first mode, and 1% in higher modes. Maximal pump depletion of 95.5% occurs at a pump power 2.52 times above threshold.

In this paper, we study the effect of various degrees of focusing for pump and signal on threshold and pump depletion for the same model. In section 1, we present the coupled wave-equations, define intrinsic units for the length scales and power, and discuss the numerical solution procedure. In section 2, we present our results for the maximal degree of pump depletion and for the required pump power for threshold and various degrees of pump depletion as functions of the confocal parameters of pump and signal. In section 3, we discuss the physical mechanisms responsible for the observed numerical trends. In section 4, we discuss the limitations of our model, and summarize our findings.

**1. Field equations, resonator modeling and numerical procedures**
**a. Field equations, intrinsic units and photon conservation**

The electric field $E_F$ and the total power $P_F$ of the pump, signal and idler fields (F=P,S,I) are given in terms of the complex field amplitude $A_F$ as:

$$E_F = \text{real}(A_F \exp(i\Phi_F)), \quad \Phi_F = k_F z - i\omega t, \quad k_F = \omega_F n_F/c = 2\pi n_F/\lambda_F \tag{1}$$

$$P_F = 0.5 n c \varepsilon_0 \int dx\, dy\, |A_F|^2$$

For steady-state continuous-wave operation, the coupled wave equations for the amplitudes $A_F$ in a second-order non-linear medium are:

$$A_{F,z} = (i/2k_F)(A_{F,xx} + A_{F,yy}) + Q_F \tag{2}$$

$$Q_P = i\,\kappa_P \exp(-i\Delta z) A_I A_S,\quad Q_S = i\,\kappa_S \exp(i\Delta z) A_P A_I^*,\quad Q_I = i\,\kappa_I \exp(i\Delta z) A_P A_S^*,$$

$$\kappa_F = \omega_F d/n_F c = 2\pi d/n_F \lambda_F,\quad \Delta = k_P - k_S - k_I - 2\pi/l.$$

Here d is the non-linear coefficient, and $\Delta = k_P - k_S - k_I - 2\pi/l$ is the residual dispersion for a periodically poled crystal with poling period l. The subscripts x, y and z denote partial differentiation with respect to the corresponding spatial coordinates. For example, for periodically poled LiNbO$_3$ (PPLN) at $\lambda_P = 1.06$ μm, one has approximately d=14 and n=2.2. To facilitate the application of our numerical data to various experimental situations, we introduce arbitrary units of propagation distance $z_0$ and wave-vector $k_0$, which define corresponding units $r_0$, $A_{F0}$ and $P_{F0}$ for the radial distance, complex amplitude and power as follows:

$$r_0^2 = z_0/k_0,\quad A_{P0} = 1/(z_0\sqrt{\kappa_S \kappa_I}),\quad A_{S0} = 1/(z_0\sqrt{\kappa_P \kappa_I}),\quad A_{I0} = 1/(z_0\sqrt{\kappa_S \kappa_P}) \quad (3)$$

With these units, the wave equations 2 and the expression for the beam power $P_F$ take the form:

$$A'_{P,z'} = (i/2k'_P)(A'_{P,x'x'} + A'_{P,y'y'}) + i\,\exp(-i\Delta' z') A'_I A'_S, \quad (4.a)$$

$$A'_{I,z'} = (i/2k'_I)(A'_{I,x'x'} + A'_{I,y'y'}) + i\,\exp(i\Delta' z') A'_P A'^{*}_S, \quad (4.b)$$

$$A'_{S,z'} = (i/2k'_S)(A'_{S,x'x'} + A'_{S,y'y'}) + i\,\exp(i\Delta' z') A'_P A'^{*}_I, \quad (4.c)$$

$$z = z' z_0,\quad k_F = k_F' k_0,\quad x = x' r_0,\quad y = y' r_0,\quad A_F = a_F' A_{F0},\quad \Delta = \Delta'/z_0.$$

$$P_F = r_F P_{0F},\quad r_F = (1/\pi) \int dx' dy'\, |A_F|^2, \quad (5)$$

$$P_{0P} = \varepsilon_0 c\, n_P n_S n_I \lambda_S \lambda_I / (8\pi\, d^2\, z_0\, k_0).$$

The units $P_{0F}$ of power are related by:

$$\lambda_P \, P_{0P} = \lambda_I \, P_{0I} = \lambda_S \, P_{0S}. \tag{6}$$

The dimensionless quantities $r_F$ are proportional to the total rates of photons passing through the crystal in unit time. They satisfy the following relationships, which are expressions of photon number conservation:

$$d(r_P + r_I)/dz' = d(r_P + r_S)/dz' = d(r_S - r_I)/dz' = 0. \tag{7}$$

In our following analysis, we set $z_0 = L/2$, where L is the crystal length, and $k_0 = k_P$. In this case, the unit $P_{0P}$ of pump power is given by:

$$P_{0P} = \varepsilon_0 \, c \, n_S \, n_I \, \lambda_S \, \lambda_I \, \lambda_P / (8\pi^2 \, d^2 \, L) = 3.352 \; 10^4 \, n_S \, n_I \, \lambda_S \, \lambda_I \, \lambda_P / (d^2 \, L), \tag{8}$$

$[\lambda] = \mu m$, $[d] = pm/V$, $[L] = mm$, $[P_{0P}] = Watt$.

The expressions in square brackets denote the units for which the above numerical expression is valid. The unit $P_{0P}$ of pump power is related to the quantity $K_3$ of reference [7] by:

$$P_{0P} = 8/((1 + \lambda_S/\lambda_P) \, K_3). \tag{9}$$

**b. Resonator modeling**

We now use the photon conservation equations 7 to derive relationships between resonator input and output in a SRO with a low-loss ring resonator. We denote the photon rates at the input and output faces of the crystal by $r_{Fin}$ and $r_{Fout}$. Then the following relationships hold:

$$r_{Iin} = 0, \quad r_{Sin} = r_S, \quad r_{Pin} = r_P, \tag{10}$$
$$r_{Iout} = r_I, \quad r_{Sout} = r_S + r_I, \quad r_{Pout} = r_P - r_I.$$

If T<<1 is the total resonator loss for the signal, one has

$$r_{Sin} = (1-T) r_{Sout} \cong r_S + r_I - Tr_S, \quad r_I = Tr_S << r_S. \quad (11)$$

In this case, the total signal power in the cavity stays almost constant. If the input pump beam is a zero order radial mode (Gaussian), the zero order radial signal mode has lowest threshold. We found that the threshold for the first radial signal mode is about four times the threshold for the zero order mode. Therefore, it is a valid approximation to replace the signal amplitude in equations 4.a and 4.b by the amplitude corresponding to a zero order mode with constant total power. Equations 4.a and 4.b are then linear in the pump and idler amplitude, and the total idler output power is a linear function of pump input power, with a conversion coefficient $D(r_S)$ depending on the signal power:

$$r_I = D(r_S) r_P. \quad (12)$$

From equations 11 and 12, the pump depletion $D_P$ and pump power are obtained as functions of $r_S$ as:

$$D_P = r_I/r_P = D(r_S), \quad (13.a)$$

$$r_P = r_I/D(r_S) = T r_S/D(r_S) = T R_P, \quad P_P = r_P P_{0P} = R_P T P_{0P}. \quad (13.b)$$

The normalized pump power $R_P$ as a function of $r_S$ depends only on the focusing geometry and ratio of idler to pump wave length, while all dependence on the resonator loss and the non linear coefficient is contained in T and $P_{0P}$. In the limit of $r_S \rightarrow 0$, equation 13.b gives the normalized threshold pump power $R_{thr}$. The gain coefficient $D(r_S)$ is calculated numerically by the procedure described below.

**c. Numerical algorithms and parameters**

Equations 4.a and 4.b were integrated numerically by a split step algorithm, with an explicit half-step of parametric conversion followed by a full step of diffraction and an implicit half step of conversion. The diffraction step uses the Cayley transform of a discrete version of the radial Laplacian which conserves photon number. This procedure is unconditionally stable [12]. The use of the radial Laplacian restricts us to radially symmetric beams, which is a reasonable approximation for many experimental situations. It results in a considerable reduction in computation time as compared to a full two-dimensional transverse integration.

As discussed above, the signal beam was set to a Gaussian with constant power $r_S$ throughout the crystal. The amplitude of the pump beam at the crystal entry face was set to be that of a Gaussian with unit power $r_P = 1.0$, while the amplitude of the idler beam was set to zero. The total output power of the idler at the crystal exit face was calculated for a sequence of values $r_S$ of signal power, to give the conversion or pump depletion coefficient $D(r_S)$, according to equation 12. At each value of $r_S$, the residual dispersion parameter $\Delta$ was adjusted to give maximal idler output power. Equations 13.a and 13.b were then used to obtain the pump depletion $D(R_P)$ as a function of the actual normalized pump input power $R_P$ corresponding to the specified values of signal power $r_S$, according to the resonator condition, equation 11.

We verified our numerical procedure extrinsic by propagating zero and first order radial modes without conversion, and plane wave beams with full conversion. Agreement of the numerical beam amplitudes with the known analytical solutions was better than 0.1 % of the maximal amplitudes, for the choice of radial and longitudinal step sizes given below. Larger errors occurred for confocal parameters smaller than about one half the crystal length at radial distances where the beam amplitudes had decreased to less than about 5% of their maximal values and longitudinal distances more than a Raleigh range from focus. These parts of the beam profile contribute less than 0.25% of the total beam power, and the errors were therefore tolerated. In [11], we also found very close agreement between the numerical results obtained from the procedure described here, and a numerical integration of coupled mode equations. Furthermore, the threshold power calculated was found to agree with the results of reference [7].

Intrinsic verification was obtained by reducing the radial and longitudinal step sizes by one half, resulting in changes of less than 0.2% of the calculated idler power, which is consistent with the results from the extrinsic verification described above. The crystal length L was defined to be two units of length: $L=2L_0$, and was divide into 80 integration steps. For the smallest pump confocal parameter $b_P=0.3L$ considered here, this gives 24 steps for the region within on Raleigh range from the focus. With a unit $k_0=k_P$ of wave vector, a pump beam with $b_P=L=2L_0$ has a waist radius of $w=1.4w_0$ units $w_0$ of radial distance. We chose the maximal radial distance to be 14 units, and the number of radial integration steps to be 200. This gives 20 steps within the waist radius for $b_P=L$, and more than 10 steps for $b_P=0.3L$.

The signal power $r_S$ was increased from $r_S=0.0001$ by 0.2 to a final value of 4.0001. Maximal conversion was found in the range of $r_S=2.8$ to 3.2. From the resulting discrete series of pump depletion or conversion coefficients the maximal conversion and pump power levels at fixed conversion were determined by interpolation.

## 2. Numerical results for pump depletion and pump power levels

Figures 1 to 9 summarize results for various parameters, which characterize the conversion efficiency of the SRO as functions of the inverse focusing parameter $z_P'=z_P/L_0=b_P/L$. Figures 1 to 6 present results for fixed signal confocal parameter $z_S'=b_S/L$, with figures 1 to 3 for the case $k_I/k_P=0.33$ and figures 4 to 6 for $k_I/k_P=0.50$. Figures 7 to 9 give corresponding results for the case where the pump and signal confocal parameters are equal: $z_P'=z_S'$.

### a. Results for $k_I/k_P=0.33$, $b_S/L=1.0$, 0.7 and 0.5.

Figure 1 shows the maximal pump depletion $D_{max}$ as a function of $z_P'$, for $z_S'=1.0$, 0.7 and 0.5. The largest value is always found at $z_P'$ smaller than or equal to $z_S'$. For example, at $z_S'=1.0$, $D_{max}$ peaks between $z_P'=0.8$ and 0.9 at about 0.965, while for $z_S'=0.5$, the peak value of 0.977 is found at $z_P'=0.5$. Thus, for maximal conversion, it is necessary to focus the pump not less tightly than the signal. The penalty for violating this requirement is not severe, however. For $z_S'=1.0$, one stays within 5% of the peak value, that is above 0.91, for $z_P'$ in the range from 0.5 to 1.5.

Figure 2 shows the pump power levels $R_P$ at which maximal depletion $D_{max}$ and

95% of maximal depletion, $D=0.95D_{max}$, are achieved. For $z_S'=1.0$, the lowest pump level required is $R_P=3.12$ for maximal depletion, and $R_P=2.34$ for 95% of maximal depletion. The corresponding value of $z_P'$ is between 0.7 and 0.8, somewhat below the value at which $D_{max}$ peaks. For $z_S'=0.5$, the minimal levels are lower: $R_P=2.87$ and 2.12, and they occur for the same $z_P'=0.5$ as for peak $D_{max}$.

Figures 3.a to 3.c show the pump levels $R_P$ required to reach threshold $D=0.0$, and the values $D=0.2$, 0.4, 0.6 and 0.8. The threshold values agree with those published in [7] (see also [10]). Minimal power levels are found in the same range for $z_P'$ as in figure 2: $z_P'=0.7$ to 0.8 for $z_S'=1.0$ (figure 3.a), $z_P'=0.6$ for $z_S'=0.7$ (figure 3.b) and $z_P'=0.5$ for $z_S'=0.5$ (figure 3.c).

Figure 3.d finally gives the ratio $P_P/P_{thr}=R_P/R_{thr}$ of pump power at $D=0.95D_{max}$ (figure 2) to threshold power (figures 3.a to 3.c). It is very interesting to note that this ratio falls into a narrow range from 1.99 to 2.07, a variation of only 4%. Thus, as a rule of thumb for a wide range of focusing conditions one expects $D=0.95D_{max}$ to occur at about 2 times above threshold. Violation of this rule will indicate the presence of nonlinear effects other than parametric conversion.

### b. Results for $k_I/k_P=0.50$, $b_S/L=1.0$, 0.7 and 0.5.

Figures 4 to 6 present corresponding results for the degenerate case $k_I/k_P=k_S/k_P=0.50$. The maximal depletion $D_{max}$ is somewhat higher than for $k_I/k_P=0.33$, although not by a large amount. For $z_S'=0.5$, it peaks just below $D_{max}=0.98$ at $z_P'=0.5$.

The pump levels are lower than for $k_I/k_P=0.33$. For example, at $z_S'=0.5$, the minimal level for $D=0.95D_{max}$ is $R_P=2.18$, while it is $R_P=2.34$ for $k_I/k_P=0.33$. The same trend is found for the relative pump power levels $P/P_{thr}$ at $D=0.95D_{max}$, shown in figure 6.d. These lie in the range from 1.93 to 1.99, as compared to the range 1.99 to 2.07 for $k_I/k_P=0.33$.

### c. Results for $b_S=b_P$, $k_I/k_P=0.33$, 0.50.

Figures 7 to 9 present results for the case where signal and pump have equal confocal parameter. Peak $D_{max}$ and lowest power levels are found for tighter focusing, around $z_P'=z_S'=0.4$. The degenerate case $k_I/k_P=0.50$ requires significantly lower pump power levels for given depletion D than $k_I/k_P=0.33$ at tighter focusing $z_P'=z_S'<1.0$, while for larger confocal parameters the difference is less. Differences in $D_{max}$ between the two cases are less than 1% for all values of $z_P'=z_S'$.

### d. Relative pump depletion as a function of relative pump power.

Figure 10 gives the relative depletion $D/D_{max}$ as a function of relative pump power $P/P_{thr}$, showing one near optimal case each for $k_I/k_P=0.33$ and 0.50, and the result of the plane wave theory. The curve for $k_I/k_P=0.50$ follows the plane wave curve more closely, staying below the plane wave curve until $D_{max}$ is reached, and declining more gradually afterwards. This behavior is more pronounced for $k_I/k_P=0.33$.

All three curves agree to better than 5% up to 4 times above threshold. Furthermore, all other cases covered in this study fall within the same range. This result shows that the influenced of various focusing configurations is mostly accounted for by the dependence of pump threshold $P_{thr}$ (figures 3,6 and 9) and maximal depletion $D_{max}$ (figures 1,4 and 7). With these two parameters, the conversion D at any pump power level P is predicted to within 5% by scaling to the plane wave formula:

$$D(P/P_{thr}) = D_{max} \, D_{pw}(P/P_{thr}) \tag{14}$$

Here $D_{pw}(P')$ is the pump depletion for plane wave beams at $\Delta'=0$ as a function of the ratio $P'=P/P_{thr}$ of pump power to threshold power. According to ref. [8], this is defined implicitly by giving $D_{pw}$ and $P'$ as functions of the signal amplitude $A_S'$ (see equations 4.a and 4.b):

$$D_{pw}(A_S') = \sin(A_S'L')^2, \quad P'=(A_S'L')^2/\sin(A_S'L')^2, \quad L'=L/z_0. \tag{15}$$

The maximal pump depletion for plane wave beams at $\Delta'=0$ is $D_{pw\,max}=1.0$.

## 3. Intensity and phase dependent effects of focusing

The dependence of SRO conversion on focusing results from both intensity and phase sensitive mechanisms.

Intensity dependence is given by the fact that tighter focusing leads to higher field amplitudes, and thereby to enhanced nonlinear conversion. This effect is counterbalanced by the reduction in effective gain length which results from stronger beam spreading. The latter effect dominates at very short confocal parameters. These combined effects are most clearly seen in figure 7 to 9, for the case $z_P'=z_S'$.

If $z_P'=z_S'$, the nonlinear polarization at the idler wave length has a radial phase dependence which, as a function of propagation distance, is that of a Gaussian with the same constant confocal parameter and location of focus as the pump and signal, as long as the pump retains its Gaussian form. Therefore, conversion can be phase matched to a zero order idler mode. On the other hand, if $z_P'$ is different from $z_S'$, the radial phase dependence of the non-linear polarization does not correspond to that of a zero order idler mode, and phase matching is frustrated. Therefore, smaller gain than in the phase matched case results, and one expects that higher pump levels are required for threshold and depletion. Also, higher transverse idler modes are expected to be generated more strongly, leading to a corresponding reduction in maximal depletion $D_{max}$, since the different transverse modes reach maximal power at different levels of signal power.

These qualitative arguments are supported by the focusing dependence of $D_{max}$ and required pump levels discussed earlier. For example, $D_{max}$ in figures 1 and 4 peaks at values of $z_P'$ close to $z_S'$. Furthermore, for $z_P'>0.7$, tighter focusing of the signal beam leads to a decrease in $D_{max}$, and for $z_P'>0.9$ also to an increase in required pump power (figures 2 and 5). This is contrary to what would be expected on the basis of intensity dependent effects only, and results form the increasing radial phase mismatch. For $z_P'<0.7$, on the other hand, tighter focusing of the signal beam increases both the signal amplitude and the degree of radial phase match, and therefore gives higher $D_{max}$ and lower required pump power.

The dependence of higher radial idler mode content on focusing is demonstrated in table 1 below, which shows the maximal depletion $D_{max}$ and the relative power in the zero order radial mode for $k_I/k_P=0.50$, $z_S'=1.0$ and $z_P'=0.4, 0.8$ and $1.5$. Best phase matching occurs for $z_P'=0.8$, which has highest $D_{max}$ and highest content of zero order radial modes. For calculation of radial mode amplitudes, the confocal parameter of the idler modes was set equal to the pump confocal parameter, and focus located at the crystal center.

| | | | |
|---|---|---|---|
| pump confocal parameter | 1.5 | 0.8 | 0.4 |
| maximal pump depletion | 0.92 | 0.97 | 0.93 |
| relative zero-mode strength | 0.65 | 0.70 | 0.62 |

Table 1: Conversion parameters for $k_I/k_P=0.5$ and $b_S/L=1.0$.

## 4. Summary and conclusion

In this paper, we presented detailed numerical results for the focusing dependence of the maximal pump depletion and of the required pump levels for various degrees of pump depletion in SROs. We found that maximal depletion occurs for values of the pump confocal parameter slightly less than or equal to the signal confocal parameter. The focusing dependence of conversion and required power levels is mostly contained in the values of pump threshold $P_{thr}$ and maximal depletion $D_{max}$. The relative depletion $D/D_{max}$ as a function of relative pump power $P/P_{thr}$ is equal to the result from the plane wave theory to within 5%. In particular, $D=0.95D_{max}$ is achieved at about 2 times above threshold.

Both intensity and phase dependent effects are important. If the confocal parameters of pump and signal are almost equal, the focusing dependence of maximal conversion and threshold is due mostly to the dependence of signal amplitude and effective gain length. For larger differences between confocal parameters, the radial phase mismatch is the dominant factor, and conversion increases with reduced phase mismatch, even if a decrease in signal amplitude is involved.

These results present a simple quantitative and qualitative account of the main effects of diffraction in SROs, which can be used to easily predict and assess the performance of focused beam based SROs. The main limitations of our model are: a) restriction to radially symmetric beams and zero order pump and signal beams, and b) neglect of any nonlinear effects other than parametric conversion and dispersion. In particular, thermal effects and resulting radial changes in refractive index may be of importance in continuous-wave SROs [13].

It is a pleasure to acknowledge interesting discussions about the topics presented here with W. Bosenberg-who stimulated our interest in SROs-, R. Byer and M. Fejer. Thanks are also due to S. Schiller for inviting us to contribute to this special issue.

### Figure Captions

Figure 1: Maximal pump depletion $D_{max}$ as a function of pump confocal parameter.

Figure 2: Required pump power for $D=D_{max}$ and $D=0.95\ D_{max}$.

Figure 3.a: Required pump power for threshold (D=0.0), D=0.2, 0.4, 0.6 and 0.8 at $b_S/L=1.0$.

Figure 3.b: Required pump power for threshold (D=0.0), D=0.2, 0.4, 0.6 and 0.8 at $b_S/L=0.7$.

Figure 3.c: Required pump power for threshold (D=0.0), D=0.2, 0.4, 0.6 and 0.8 at $b_S/L=0.5$.

Figure 3.d: Required relative pump power $P/P_{thr}$ for $D=0.95\,D_{max}$.

Figure 4: Maximal pump depletion $D_{max}$ as a function of pump confocal parameter.

Figure 5: Required pump power for $D=D_{max}$ and $D=0.95\,D_{max}$.

Figure 6.a: Required pump power for threshold (D=0.0), D=0.2, 0.4, 0.6 and 0.8 at $b_S/L=1.0$.

Figure 6.b: Required pump power for threshold (D=0.0), D=0.2, 0.4, 0.6 and 0.8 at $b_S/L=0.7$.

Figure 6.c: Required pump power for threshold (D=0.0), D=0.2, 0.4, 0.6 and 0.8 at $b_S/L=0.5$.

Figure 6.d: Required relative pump power $P/P_{thr}$ for $D=0.95\,D_{max}$.

Figure 7: Maximal pump depletion $D_{max}$ as a function of pump confocal parameter for $b_P=b_S$.

Figure 8: Required pump power for $D=D_{max}$ and $D=0.95\,D_{max}$, for $b_P=b_S$.

Figure 9.a: Required pump power for threshold (D=0.0), D=0.2, 0.4, 0.6 and 0.8 for $b_P=b_S$ at $k_I/k_P=0.33$.

Figure 9.b: Required pump power for threshold (D=0.0), D=0.2, 0.4, 0.6 and 0.8 for $b_P=b_S$ at $k_I/k_P=0.50$.

Figure 9.c: Required relative pump power $P/P_{thr}$ for $D=0.95\,D_{max}$.

Figure 10: Relative pump depletion $D/D_{max}$ as a function of relative pump power $P/P_{thr}$.

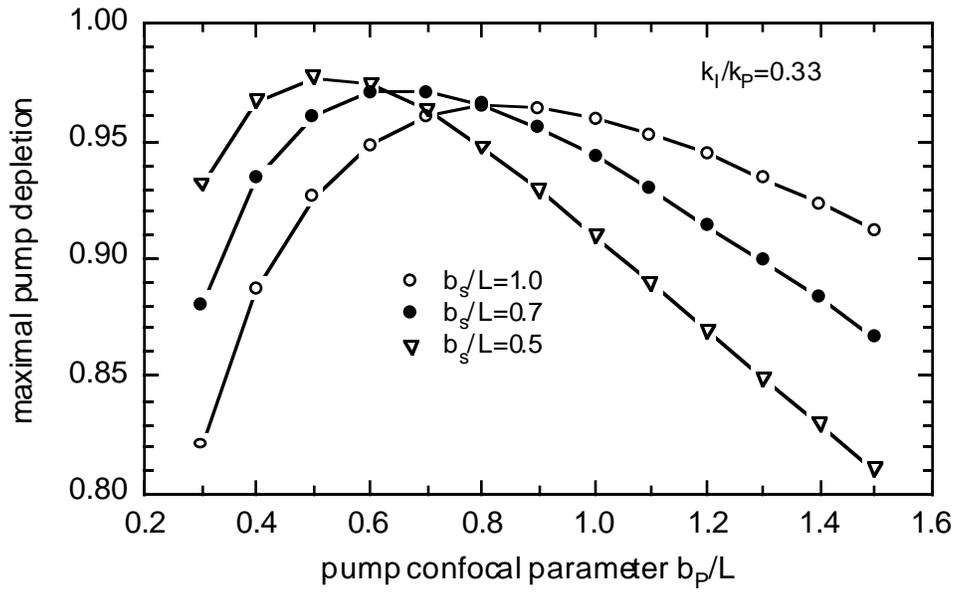

Figure 1. Maximal pump depletion

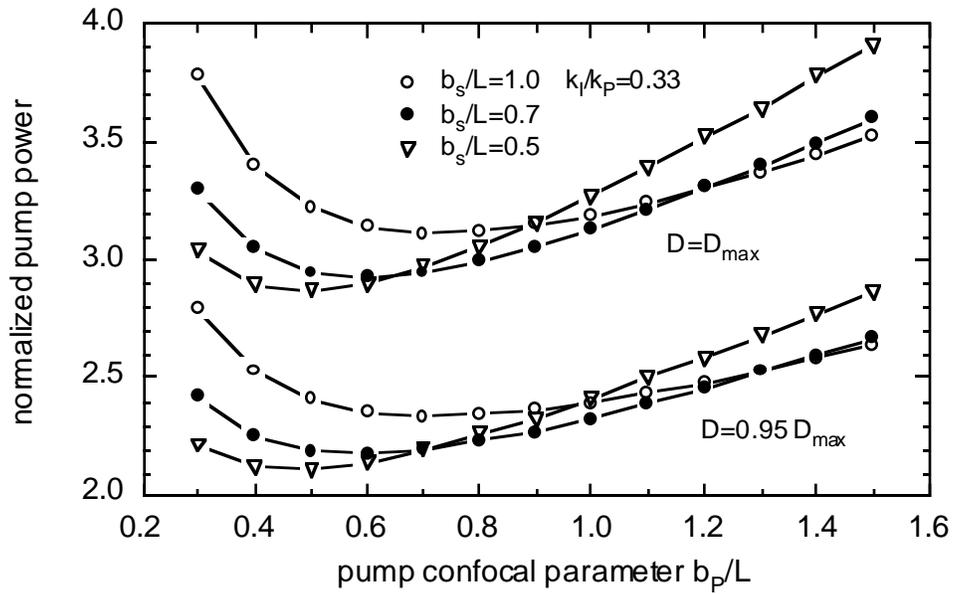

Figure 2: Pump power at maximal depletion

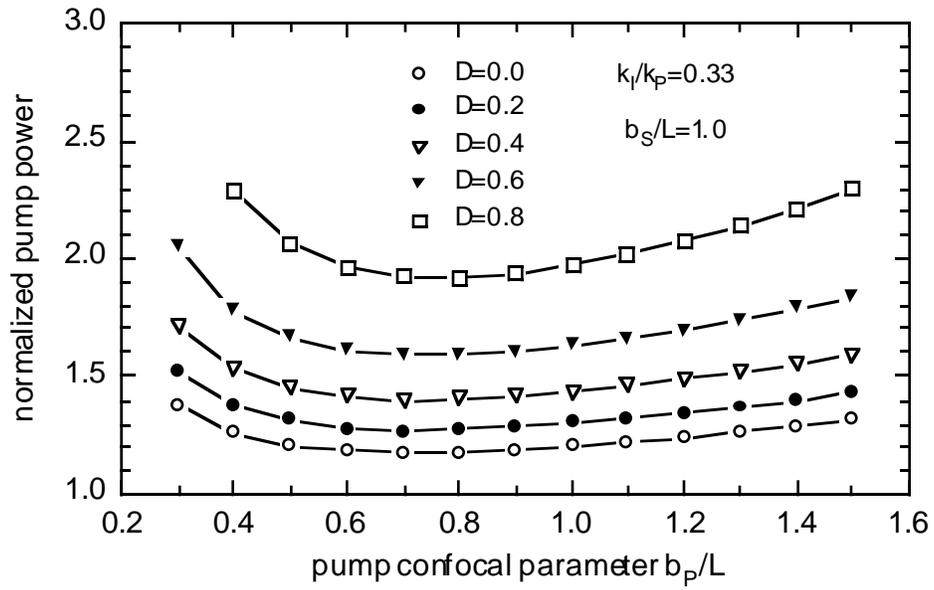

Figure 3.a: Pump power at constant depletion

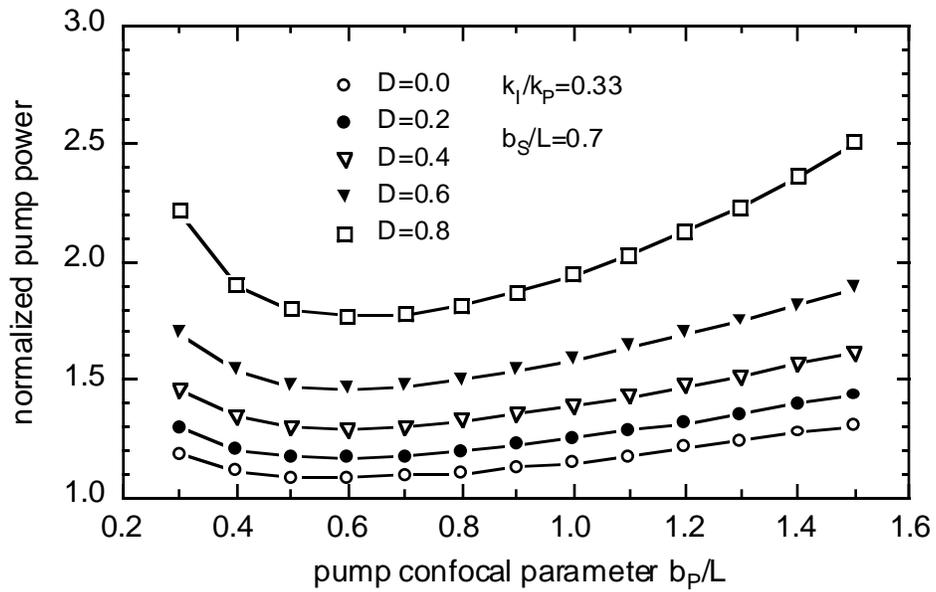

Figure 3.b: Pump power at constant depletion

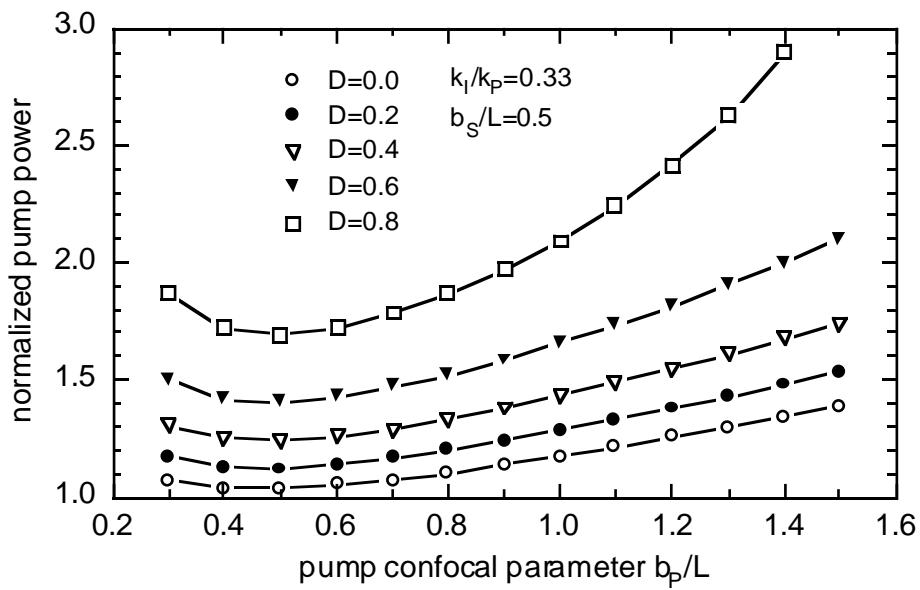

Figure 3.c: Pump power at constant depletion

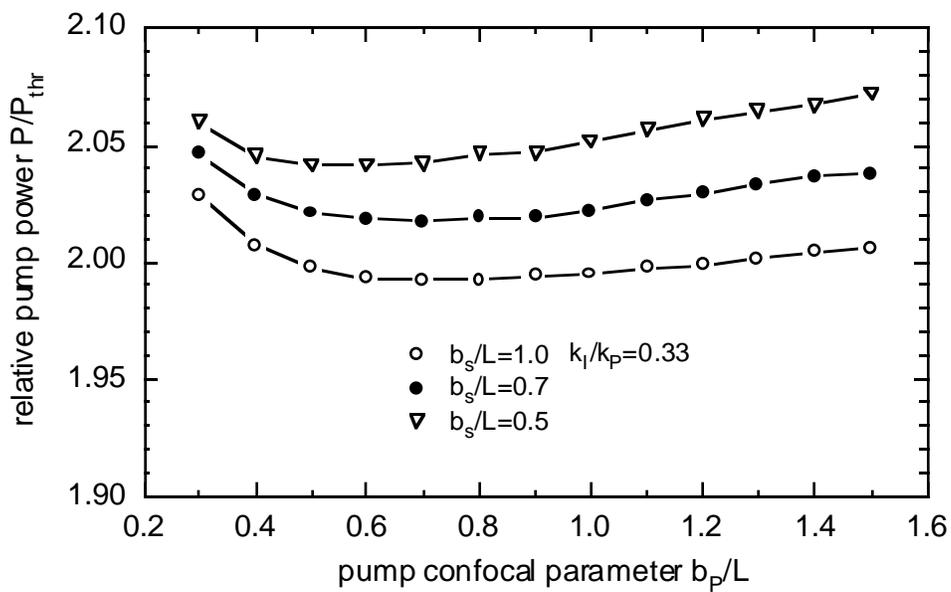

Figure 3.d: Pump power at $D=0.95\,D_{max}$

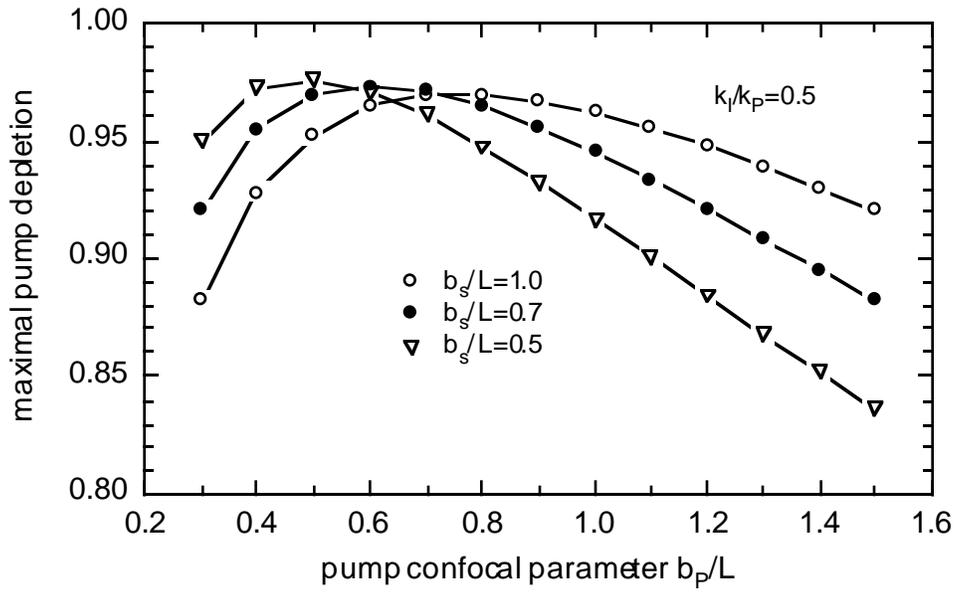

Figure 4. Maximal pump depletion

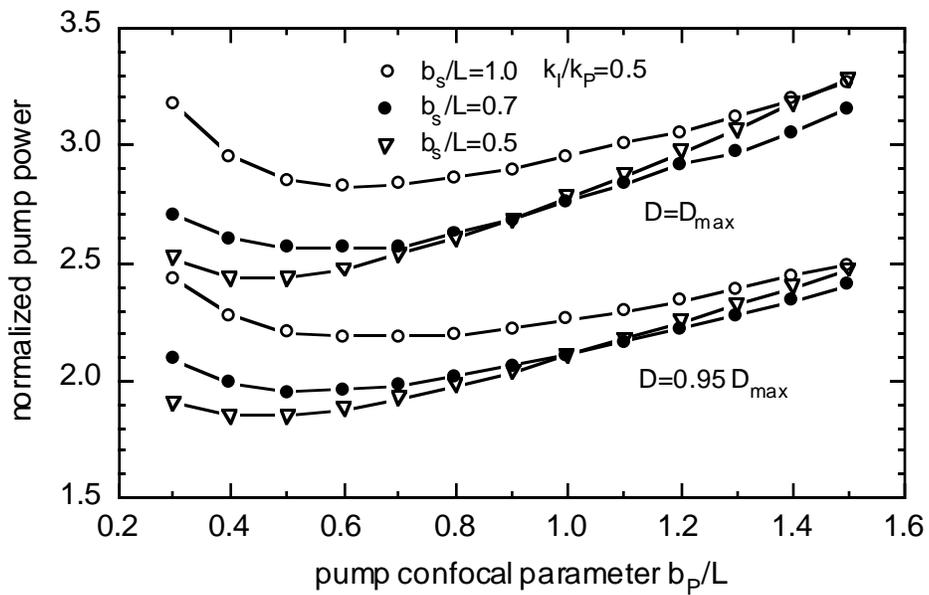

Figure 5: Pump power at maximal depletion

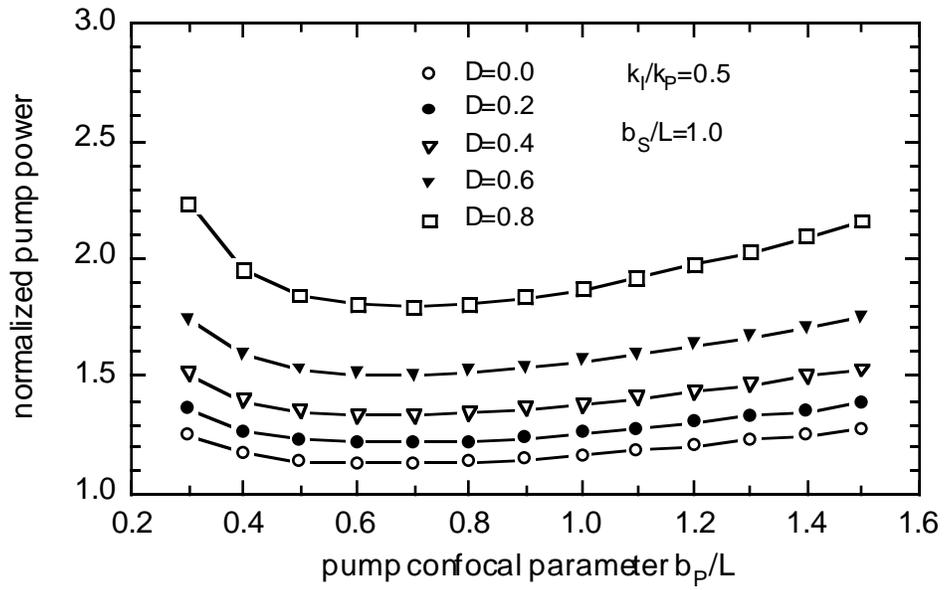

Figure 6.a: Pump power at constant depletion

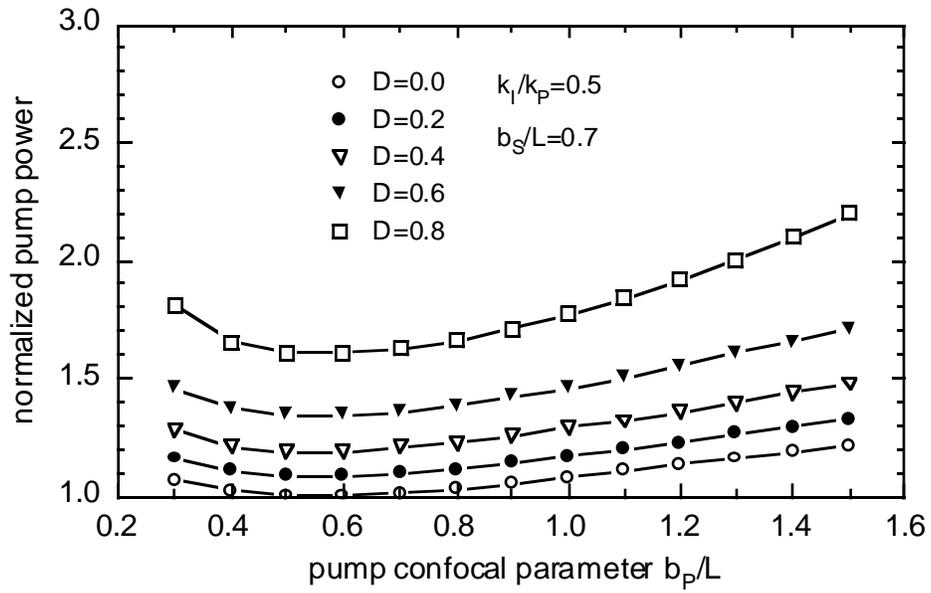

Figure 6.b: Pump power at constant depletion

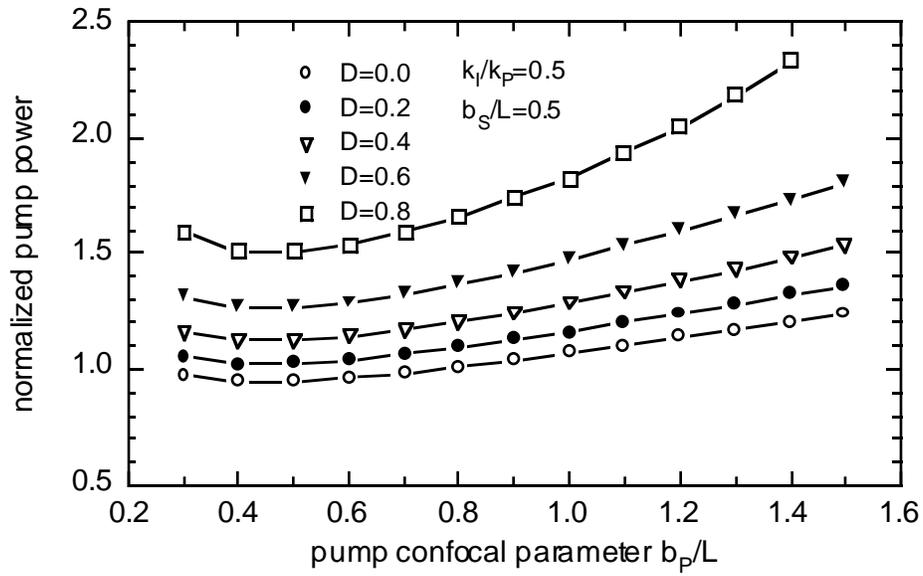

Figure 6.c: Pump power at constant depletion

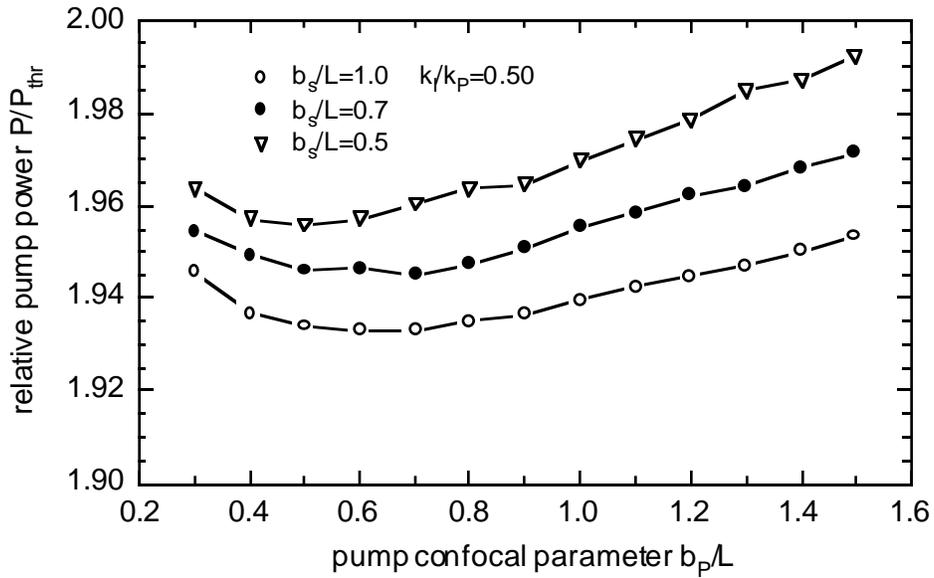

Figure 6.d: Pump power at D=0.95 $D_{max}$

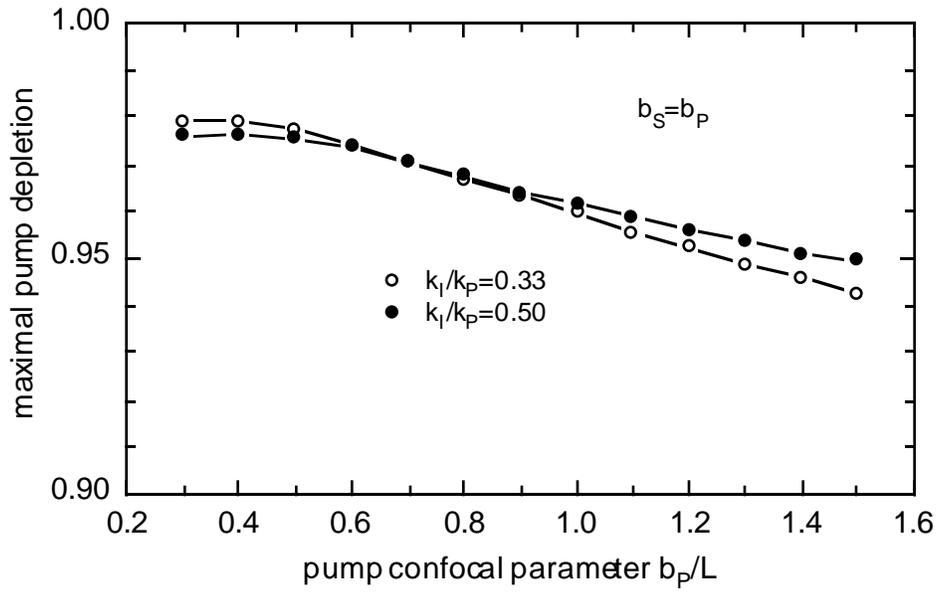

Figure 7. Maximal pump depletion

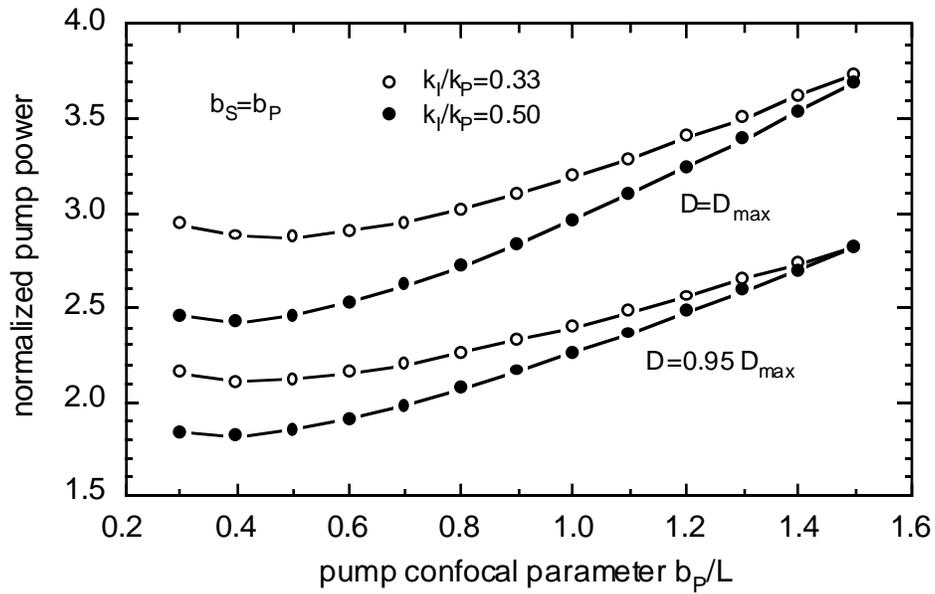

Figure 8: Pump power at maximal depletion

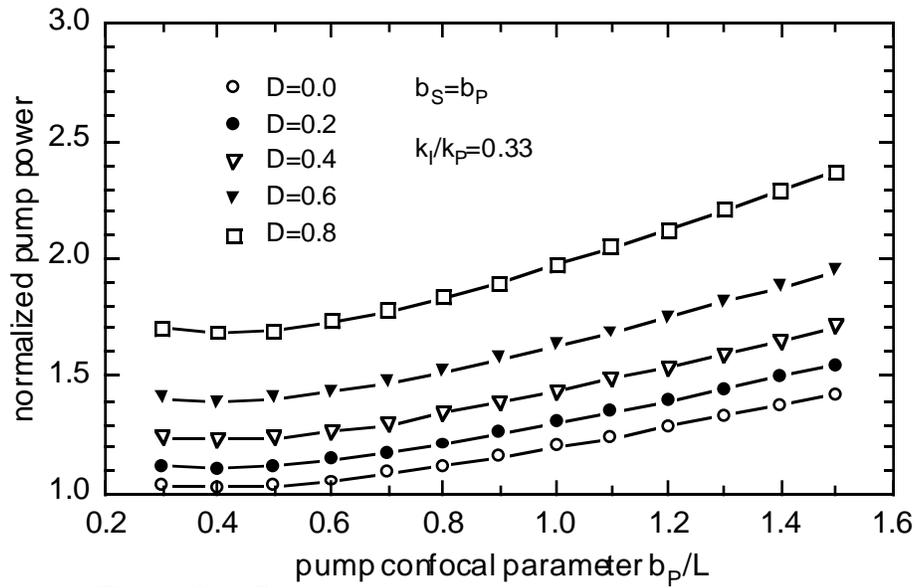
Figure 9.a: Pump power at constant depletion

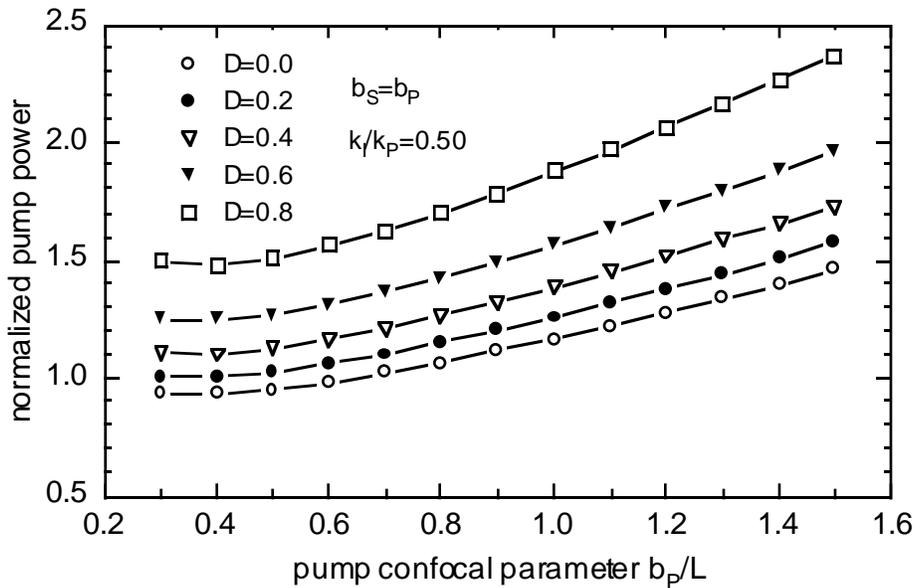
Figure 9.b: Pump power at constant depletion

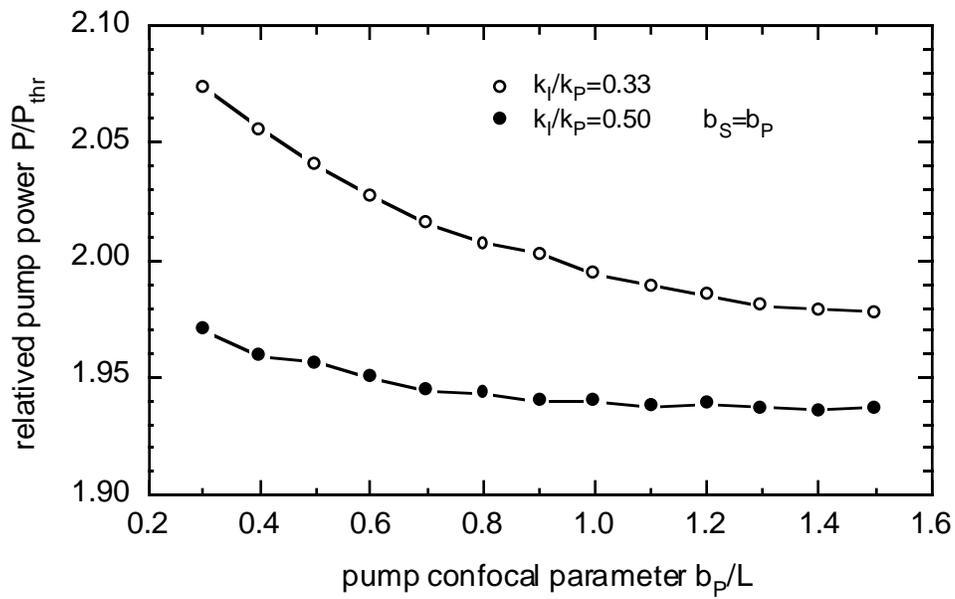

Figure 9.c: Pump power at $D=0.95\,D_{max}$

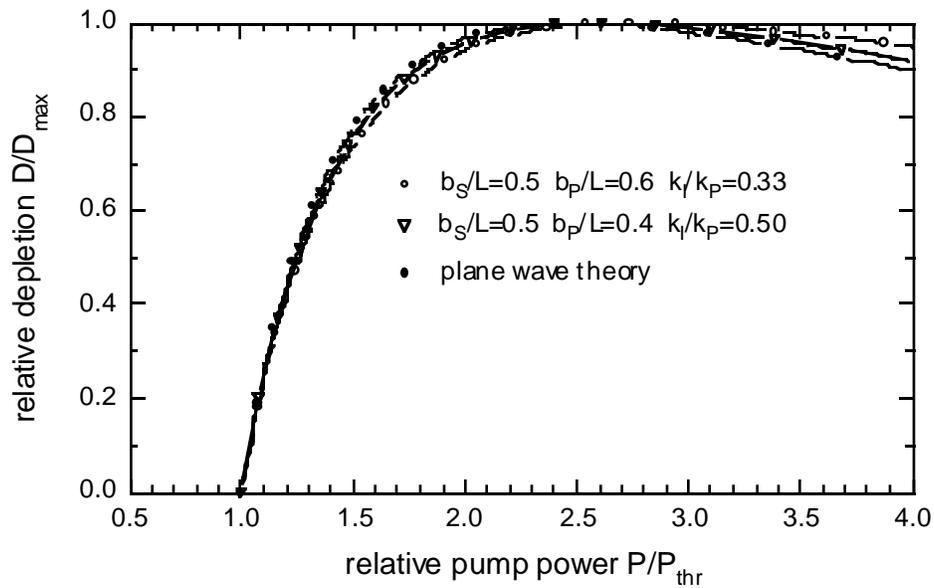

Figure 10: Relative pump depletion $D/D_{max}$